\documentclass[apj,letter]{emulateapj}
\usepackage{graphicx}

\makeatletter

\newcommand{\noun}[1]{\textsc{#1}}
\providecommand{\tabularnewline}{\\}

\makeatother

\begin{document}

\title{Initial Conditions For Large Cosmological Simulations}

\author{S. Prunet,\altaffilmark{1} C. Pichon,\altaffilmark{1,4} D. Aubert,\altaffilmark{2,1} D. Pogosyan\altaffilmark{3},
R. Teyssier,\altaffilmark{4} \& S. Gottloeber\altaffilmark{5} }

\altaffiltext{1}{Institut d'Astrophysique de Paris, UMR 7095, 98bis
Boulevard Arago, 75014 Paris, France}
\altaffiltext{2}{Observatoire Astronomique de Strasbourg, UMR 7550,
11 rue de l'Universite, 67000 Strasbourg, France}
\altaffiltext{3}{Department of Physics, University of Alberta, Edmonton,
Alberta, T6G 2G7, Canada}
\altaffiltext{4}{CEA/DAPNIA/SAP, l'Orme des Merisiers, 91170, Gif sur
Yvette, France}
\altaffiltext{5}{AIP, An der Sternwarte 16, 14 482, Potsdam, Germany}

\begin{abstract}
This technical paper describes a software package that was designed
to produce initial conditions for large cosmological simulations in
the context of the \noun{Horizon} collaboration. These tools generalize
E. Bertschinger's \texttt{Grafic1} software to distributed parallel
architectures and offer a flexible alternative to the \texttt{Grafic2}
software for {}``zoom'' initial conditions, at the price of large
cumulated cpu and memory usage. The codes have been validated up to
resolutions of $4096^{3}$ and were used to generate the initial conditions
of large hydrodynamical and dark matter simulations. They also provide
means to generate constrained realisations for the purpose of generating
initial conditions compatible with, e.g. the local group, or the SDSS
catalog.
\end{abstract}


\section{Introduction}

\label{s:Intro} Numerical simulations have  proved to be valuable
tools in the field of cosmology and galaxy formation. They provide
a mean to test theoretical assumptions, to predict the properties
of large scale structures (and galaxies within) and give access to
synthetic observations without sacrifying the whole complexity that
arise from non-linearities. Thanks to the recent progresses in terms
of numerical techniques and available hardware, numerical cosmology
has become one of the most important (and CPU consumming) field among
the scientific topics that require extreme computing. Over the last
few years, a series of large simulations have been produced by, among
others, \citet{2000ApJ...538...83C}, the Virgo Consortium (\citealt{2000astro.ph..7362F},
the Millenium: \citealt{2005Natur.435..629S}), \citet{2002ApJ...571...15W},
the Gasoline team (\citealt{2004NewA....9..137W}). Following the
same route, the purpose of the \noun{Horizon} Project%
\footnote{http://www.projet-horizon.fr%
} is to federate numerical simulations activities within the french
comunity on topics such as : the large scale structure formation in
a cosmological framework, the formation of galaxies and the prediction
of its observational signatures. The collaboration studies the influence
on the predictions of the resolution, the numerical codes, the self-consistent
treatment of the baryons and of the physics included.

These investigations are performed on initial conditions (ICs
thereafter) that share the same phases and their production is
described in the current paper.  The large \noun{Horizon} ICs involves
two boxes of 50 and 2000 Mpc/h comoving size with respectively
$1024^{3}$ and $4096^{3}$ initial resolution elements (particles or
grid points), following a $\Lambda$CDM concordance cosmogony. They
were generated from an existing set of initial conditions created for
the 'Mare Nostrum' simulation (\citealt{2007ApJ...664..117G}): they
share the same phases but with different box sizes and
resolutions. The ($50h^{-1}$Mpc,$1024^{3}$) ICs were used as inputs to
the AMR code RAMSES (\citealt{2002A&A...385..337T}) in a simulation
that included dark matter dynamics, hydrodynamics, star formation,
metal enrichment of the gas and feedback. This simulation directly
compares to the Mare-Nostrum simulation in terms of cosmology and
physics and it will be refered as the \noun{Horizon}-MareNostrum
simulation hereafter(\citealt{Ocvirk2008}). The
($2000h^{-1}$Mpc,$4096^{3}$) ICs served as a starting point for the
\noun{Horizon}-4$\Pi$
simulation(\citealt{Teyssier2007}, \url{http://www.projet-horizon.fr}):
it is a pure dark matter simulation and assumes a cosmology
constrained by WMAP3. It is currently used to investigate the
full-sky gravitational lensing signal that could be observed by the DUNE
experiment (hence the $4\Pi$).

The paper is organised as follows: first, we briefly explain the
principle of the ICs' generation. Then we describe how the phases were
extracted from the MareNostrum ICs in order to make the \noun{Horizon}
ICs consistent with this reference. We describe next the features of a
series of codes used to generate and process the different
\noun{Horizon} ICs:
\begin{itemize}
\item \texttt{mpgrafic}: ICs generation with optional low-frequency constraints
\item \texttt{constrfield}: Low-frequency ICs generation with
  point-like constraints.
\item \texttt{degraf}: Low-pass filtering and resampling of ICs
\item \texttt{powergrid}: ICs empirical power spectrum estimation.
\item \texttt{splitgrafic}: Estimation of matter density on a grid
  from a set of particle positions, and Peano-Hilbert domain
  decomposition.
\end{itemize}
Finally we illustrate how these codes were implemented on
the two \noun{Horizon} simulations.

\section{Random Field for Cosmological Initial Conditions}

\subsection{Grid-based initial conditions}

\label{s:gen} For completeness, we quickly review the principle of
ICs generation. Most of the following has been strongly inspired from
articles by \citet{pen} and \citet{grafic2}. Let us consider the
initial 3D gaussian random field $\delta({\bf x})$, representing
the density or the displacements, and let us define its Fourier transform
$\delta({\bf k})$. If we consider zero-mean fields, they are completely
defined by their correlation function or, equivalently, by their power
spectra, $P(k)$: \begin{equation}
P(k)\delta_{D}({\bf k}-{\bf k'})=\langle\delta({\bf k)\delta^{*}({\bf k')\rangle.}}\end{equation}
 All the statistical information in a gaussian homogeneous and isotropic
realization is contained in this quantity and the difficulty of generating
initial conditions resides in obtaining a field which has the
correct power spectrum.\\
 We chose to follow the convolution-based method described by e.g
\citet{pen} and define the correlation kernel in Fourier space as:
\begin{equation}
A(k)=\sqrt{P(k)}.\end{equation}
To reproduce the correlation function accurately one may need to first
convolve the power spectrum with the window that describes the simulation
box, as advocated by \citet{pen}. The influence of the box size on
rms density in spheres of given radius (which are relevant for mass
function estimates of collapsed objects), is negligible for sphere
radii much smaller than the box length, even for simulations designed
to study galaxy cluster scales.

Then the ICs generation is a two-step procedure. First a normal, uncorrelated
random field of unit variance is generated in position
space\footnote{We could have directly generated the real and imaginary
  part of each $\delta({\bf k})$ following a
  $\mathcal{N}(0,1/\sqrt{2})$ law, saving the cost of an extra Fourier
  transform, but we chose to remain compatible with the
  \texttt{Grafic} code.}. This
\textit{white noise} $n_{1}({\bf x})$ has a constant power spectrum,
e.g.: \begin{equation}
\langle|n_{1}({\bf k)}|^{2}\rangle=1.\end{equation}
 The second step involves convolving the white noise with the correlation
kernel in order to obtain the initial fluctuation field: \begin{equation}
\delta({\bf x})=n_{1}({\bf x})*A({\bf x}),\end{equation}
 or in Fourier space \begin{equation}
\delta({\bf k})=n_{1}({\bf k})A({\bf k}).\label{e:master}\end{equation}
 It can be easily seen that $\langle|\delta({\bf k})^2|\rangle{}=P(k)$
and the initial field automatically has the correct power spectrum.

One of the main virtue of the method resides in the possibility of
using the same white noise for different power spectra. In other words,
it decouples explicitely the phases (which contain the specificities
of a given realization in terms of relative positions) from the amplitudes
of the fluctuations (corresponding to one's favorite choice of cosmological
model). A change in the physics or in the box size results in a change
of the convolution kernel, but the underlying structure of the field
will remain globally the same for a given white noise realization.
Conversely, Eq. \ref{e:master} implies that the initial phases can
be recovered from a set of ICs, provided that the convolution operation
can be inverted. In other words, it is possible to generate a new
set of ICs from an old one (see Section \ref{sec:Application-to-large})
and such a set would share the same overall structures with e.g. a
modified cosmology or box size.

\subsection{Grid-based versus particle-based initial conditions}

In numerical simulations, the dark matter distribution is almost exclusively
described in terms of particles and this discretized description is
also applied to the gas in SPH-like hydrodynamical codes. Consequently,
dealing with 'particle-type' data is the most frequent case while
the current procedure naturally deals with densities and velocities
sampled on a grid. This can be easily tackled by recalling the density-velocity
relation that is valid in the linear regime~: \begin{equation}
\frac{1}{aH}\nabla_{{\bf x}}{\bf u}=-f(\Omega_{m},\Omega_{\Lambda})\delta({\bf x}).\label{e:bernard}\end{equation}
 Here ${\bf u}$ stands for the comoving peculiar velocity, ${\bf x}$
 for the comoving position, $H$ for the Hubble
constant, $a$ for the scale factor and $f$ is defined as the logarithmic
time derivative of the growth factor: \begin{equation}
f(\Omega_{m},\Omega_{\Lambda})\equiv\frac{{\mathrm{d}}\log D^{+}}{{\mathrm{d}}\log a}.\label{eq:deff}\end{equation}
\label{sub:In-numerical-simulations,} Functional fits for $f$ can
be found in the literature (e.g. Lahav et al. 1991) or be directly
computed for a given cosmology. Hence, assuming that particles were
displaced from a regular grid and knowing their velocities, the initial
density field can be directly recovered from Eq. (\ref{e:bernard}).
One can see that the (eulerian) positions of particles are not directly
involved in this procedure; however, their lagrangian coordinates
are used to remap the particle velocities to grid cells.

\section{Grid-based ICs: Technical Implementation}

Once the {}``phases'' (the white noise) are chosen on a grid of
a given size, it is possible to use them to generate initial density
and velocity realizations with the desired cosmology and power spectrum,
at resolutions that can differ from the initial white noise realization.

 According to linear theory, which applies for initial conditions
\footnote{actually, the validity of the linear theory is enforced by chosing the starting
redshift in such a way that the resulting variance of the discrete
density field is significantly less than unity%
}, density and velocity divergence are related through Eq.~(\ref{e:bernard}),
so that a single white noise realization determines both density and
velocities on a grid of equal size (see e.g. \citealt{grafic2}).\\
 A first numerical implementation of this algorithm was made by E.
Bertschinger in the package \texttt{Grafic1}. We extend here his code
using the Message Passing Interface (MPI) library to deal with large
simulation cubes on distributed memory platforms. Another implementation
of MPI-based Initial Conditions generator in the context of the GRACOS
code is described in \citet{2007arXiv0711.4655S}, \url{http://www.gracos.org}.\\
 We also develop a few tools (low-pass filtering and resampling, power
spectrum estimation, estimation of matter density on a grid from a
set of particle positions) that work as well in parallel environments.
We describe these tools and their usage in the following subsections.

\begin{figure*}[t]
\begin{centering}
\includegraphics[width=0.6\columnwidth]{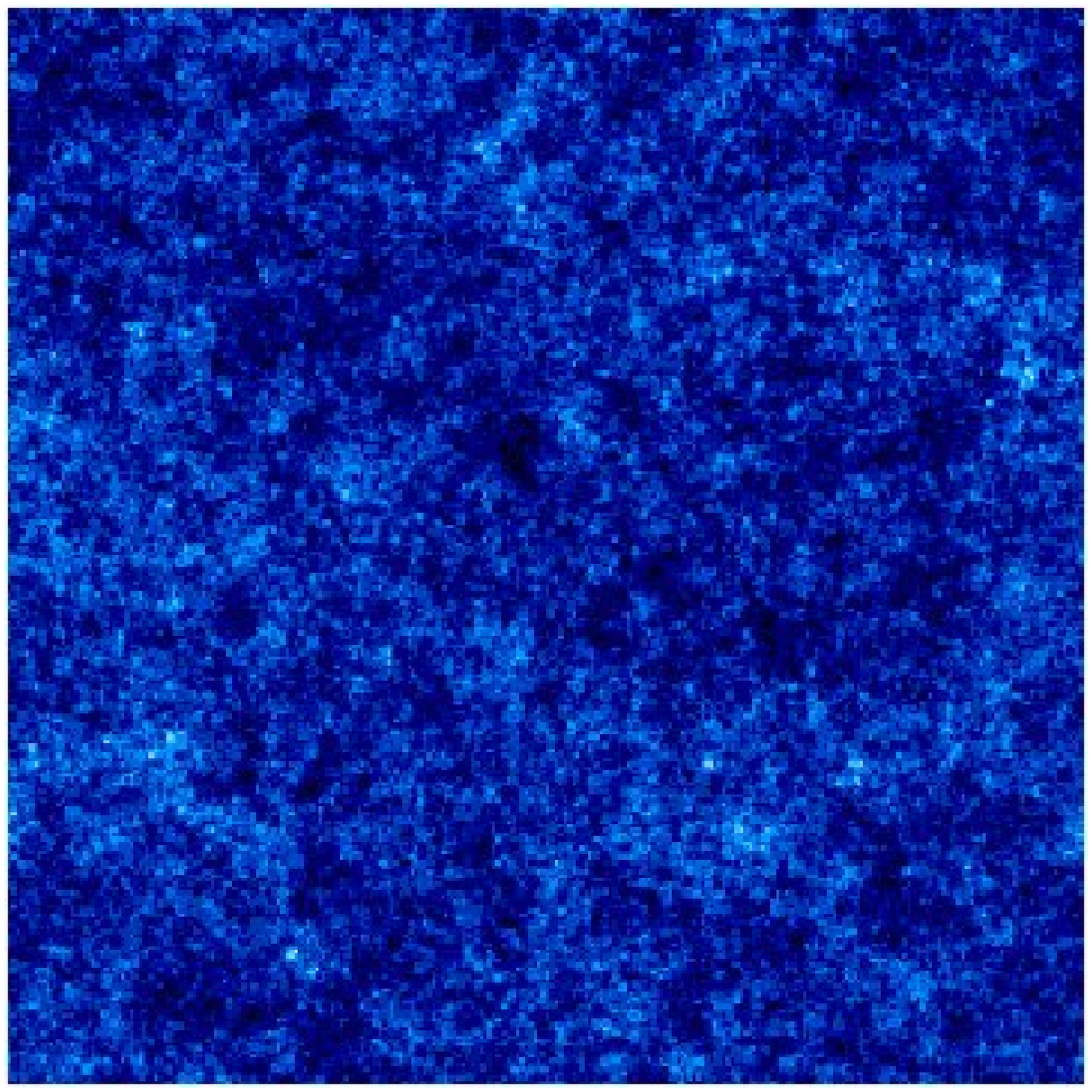} \includegraphics[width=0.6\columnwidth]{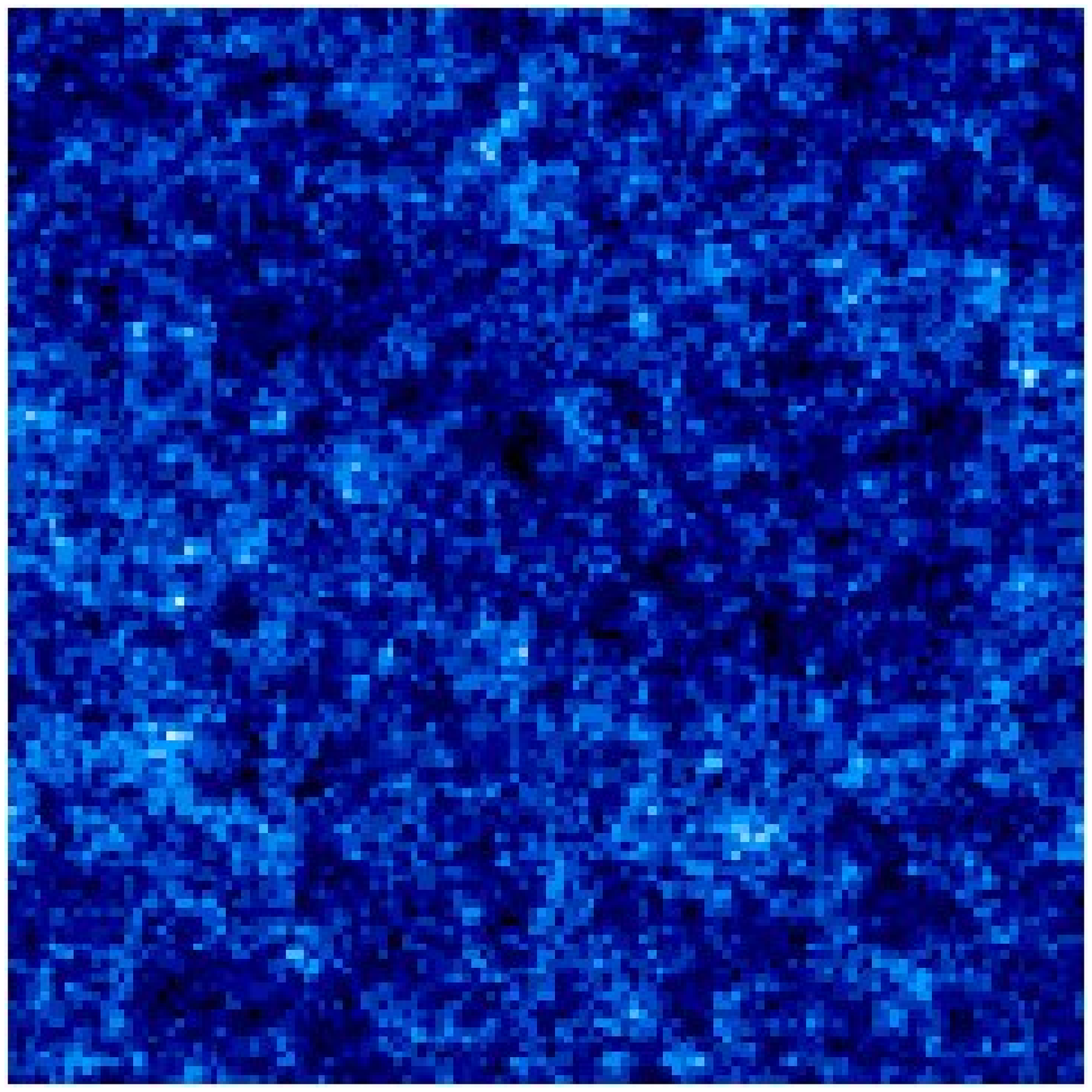}
\includegraphics[width=0.6\columnwidth]{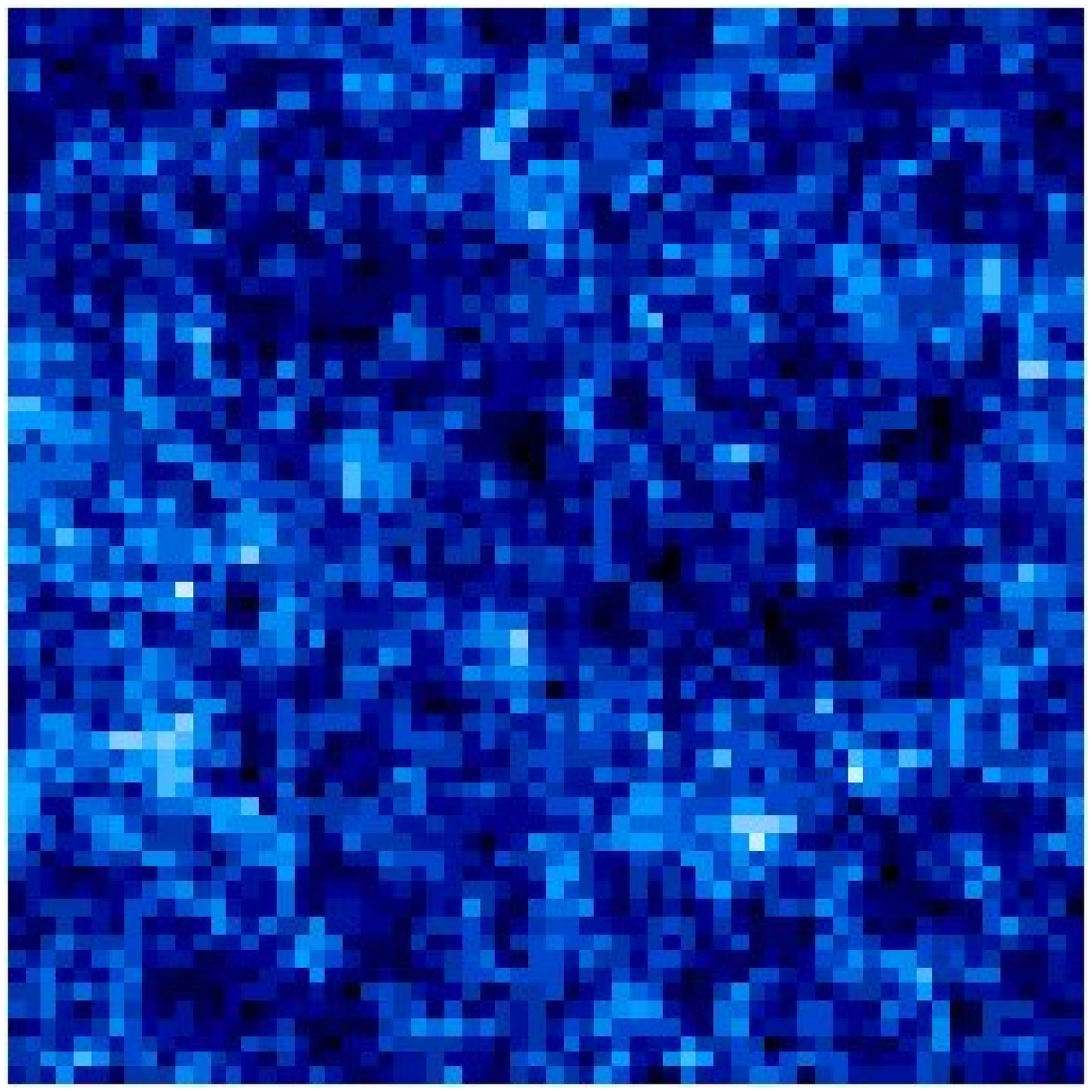}
\par\end{centering}

\caption{The left image shows a slice of a density field realization of size
$256^{3}$. The middle and right images show respectively slices of
the cubes of size $128^{3}$and $64^{3}$obtained from the first density
field by lowpass filtering and subsampling. The images have been rescaled
to the same size to ease their comparison. The initial density field
was obtained using \texttt{mpgrafic} with the parameters of Table~\ref{tab:paramfile}.
The low-frequency, resampled fields were obtained with the \texttt{degraf}
utility.\label{fig:mpdegraf}}

\end{figure*}

\subsection{\texttt{Mpgrafic}: a parallel version of \texttt{Grafic1}}

Generating initial conditions for cosmological simulations on a grid
from an initial white noise realization is theoretically quite simple,
as it involves a straightforward implementation of Equations~(\ref{e:master})
and (\ref{e:bernard}) in Fourier space. Indeed, the main issues in
the \texttt{Grafic1} code involve getting the normalization right
(in terms of $\sigma_{8}$ or $Q_{{\rm rms}}$, which are input parameters),
and therefore in the cosmology routines.

\subsubsection{Description}

Difficulties of a more technical nature appear as the size of the
desired cubes (and/or the number of particles) grow and that a single
cube does not fit into a computer's memory. An elegant answer to this
problem, in the context of multi-resolution ({}``zoom'') initial
conditions was designed by \citet{grafic2} and implemented in the
\texttt{Grafic2} package. This solution involves generating a low
resolution cube first, and successively adding higher frequencies
in nested sub-regions, constrained by the phases of the already existing
low frequency modes. Strictly speaking, the exact solution to this
problem is (naively) as costly as the direct generation of the full
cube at the highest resolution, but approximate, less costly solutions
based on anti-aliasing kernels can be designed. This is precisely
what is done in the \texttt{Grafic2} package.

The main advantage of this solution is to produce multi-resolution
initial conditions with a very reasonable memory usage, but it also
has drawbacks, namely its complexity, and its built-in restrictions
on the nested cubes structure of a given maximum size. Given the growing
size of computer clusters, our {}``brute force'' approach to the
problem based on parallelism becomes possible, and it is also in some
ways more flexible. First of all, it allows for a direct generation
of global initial conditions for very large cosmological simulations.
Secondly, it also allows, together with associated tools for low-pass
filtering and resampling, to create multi-resolution simulations of
a more general structure by simply extracting the desired sub-regions
from the series of {}``downgraded'' cubes (obtained by low-pass
filtering and resampling of the initial large high-resolution cube).

There are two issues that arise when implementating a parallel (MPI-based)
version of the \texttt{Grafic1} software. The first involves performing
efficient three-dimensional fast fourier transforms (FFTs thereafter);
this difficulty is solved by using the parallel version of the FFTW
\footnote{\texttt{http://www.fftw.org}%
} library, which uses a slab domain decomposition. The second difficulty
lies in the input/output. Indeed, we decided to keep the binary (Fortran)
structure of the files in the \texttt{Grafic1} format. In a parallel
environment where each MPI process is responsible for one chunk of
data, this lead us to write part of the I/O subroutines in C using
reentrant read/write routines, wrapped in fortran90 to be callable
from the main program.

Finally, apart from the parallelism of \texttt{mpgrafic}, we have
added a few new features to the original \texttt{Grafic1} code. An
implementation of the matter power spectrum with baryon oscillations
was introduced, as described in \citet{eisenstein}. Secondly, the
possibility of constraining the low frequency phases of the density
and velocity realizations was implemented, with the input of a given
white noise cube of lower resolution. This allowed us to use the same
set of initial condition phases for cosmological simulations at different
resolutions. Lastly, we have implemented the possibility of constraining
the value of the density or velocity field values, as well as their
gradients and hessians at a chosen set of positions. We will come
back in more details on this last point in a following section, as
it is a non-trivial extension of the code.

\subsubsection{Code installation and parameter file structure}

A prerequisite to use \texttt{mpgrafic} is to have (of course) a valid
MPI library installed, including a fortran90 compiler. A second prerequisite
is to have the \texttt{fftw-2.1.5} library installed, with the \texttt{-{}-enable-mpi
-{}-enable-type-prefix} options at the configure step. The first option
builds the static and shared FFTW MPI-based libraries, the second
is for the type (float or double) naming scheme of the libraries.

 Note that the default build of FFTW corresponds to double precision,
whereas the default type in \texttt{mpgrafic} is single precision.
To change to single precision realizations, you need to make the single
precision FFTW MPI libraries by adding the \texttt{-{}-enable-float}
at configure time. To compile \texttt{mpgrafic} in double precision
mode, you need to configure with the \texttt{-{}-enable-double} keyword.

The code usage has been kept as close as possible to the \texttt{Grafic1}
code interface, except of course for the few additional options to
the code. In table~\ref{tab:paramfile} we show an example of parameter
file of \texttt{mpgrafic}.

Compared to the original \texttt{Grafic1} parameter files, the only
differences lie in the possibility of an input power spectrum with
baryonic oscillations (\citealt{eisenstein}), and in the optional
input of a small noise file to constrain the large scale phases. Otherwise,
the code is called in the following way:\\
 \texttt{mpirun -np $<\#$proc$>$ mpgrafic $<$ parameter\_file}\\
 Like \texttt{Grafic1}, it produces seven data cubes (one density
file, three dark matter velocity files, and three baryon velocity
files). In the file \texttt{grafic1.inc}, the offsets of the velocity
fields is controlled by the parameters \texttt{offvelb, offvelc}.
The redshift of the realizations is controlled by the variance of
the density field on the grid, as specified by \texttt{sigstart}.
Finally, the size of the cubes is controlled by the parameters \texttt{np1,np2,np3},
that can take different values and are set in \texttt{grafic1.inc}.
Note that when these parameters are changed the code needs recompilation.

\subsubsection{Illustration on a small example, power spectrum estimate}

In figure~\ref{fig:mpdegraf} we show a slice of a density file
realization with the parameter file displayed in
Table~\ref{tab:paramfile}, but without large scale constraints, and
for \texttt{np1}=\texttt{np}2=\texttt{np}3=$256$.  In the same figure,
we show the density files of linear grid size $128$ and $64$ obtained
from this realization, by low-pass filtering and subsampling. This
operation has been done using the utility called \texttt{degraf}, that
makes use of the FFTW parallel Fourier transform routines. Written in
fortran90, it takes as input a collection of files in \texttt{Grafic1}
format, as well as some parameters in a namelist file. These
parameters include the list of the input file names, the target
resolution of the output files, as well as an optional shift vector
allowing a global translation of the output files (with periodic
initial conditions).

\begin{figure}[tbh]
\begin{centering}
\includegraphics[width=0.9\columnwidth]{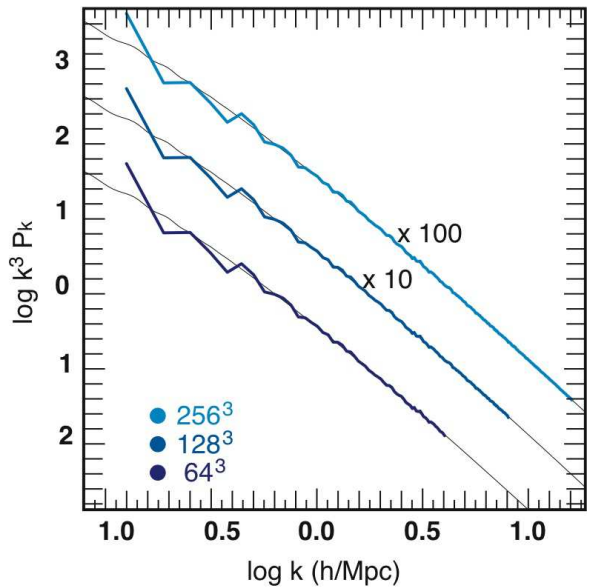}
\par\end{centering}
\caption{Theoretical power spectrum (solid line, as output in the file power.dat
by \texttt{mpgrafic}), together with the periodogram estimates (computed
by powergrid), of the $256^{3}$density realization (top thick line),
and its downgraded versions produced by degraf (middle and bottom
lines for $128^{3}$and $64^{3}$ respectively).\label{fig:powergrid} }
\end{figure}

Finally, another utility, \texttt{powergrid}, uses the FFTW parallel
Fourier transform routines to compute the periodogram estimate of
a density field power spectrum. It allows correction for nearest grid
point (NGP) or cloud in cell (CIC, linear) interpolation of a particles
set to the computation grid. Note that this resampling of a discrete
density field given by particles onto grid cells leads not only to
a smoothing of the grid Fourier modes (this can be corrected by the
code) but also to some power aliasing, that cannot be corrected, unless
prior knowledge of the density power spectrum is available. These
points will be illustrated in the next section, for the \noun{Horizon}
simulations. Of course, none of these problems appear if one is only
interested in computing the periodogram of the IC grid-sampled density
fields. Figure~\ref{fig:powergrid} shows the theoretical power spectrum
corresponding to the density field realization shown in Figure~\ref{fig:mpdegraf},
together with its periodogram estimate, as well as the periodograms
of its low-passed, resampled versions.

\subsection{Constrained initial conditions}

Since \texttt{mpgrafic} opens the opportunity of generating ICs which
are consistent with a given low frequency cube, it is interesting to
build such a cube of phases so that the overall cube satisfy (low
frequency) point like constraints on a given set of points. These
constraints fix the mean value of the density or velocity field, as
well as their gradients and hessians, computed at a chosen set of
positions, for a given smoothing length $R_p$ (see
Equation~\ref{eq:filtered-constraint} below). 
Once such a low-frequency cube has been generated, it is then
whitened, and set as an
input to \texttt{mpgrafic}, which is then used to add small scale
power and to resample the field according to the
new Nyquist frequency \footnote{In principle, adding small scale power
  with \texttt{mpgrafic}, while keeping the same low-frequency phases, 
  violates the point-like constraints imposed by
  \texttt{constrfield}. However, if the smoothing kernel $W(kR_p)$ and
  its associated smoothing length are chosen in such a way that it
  cuts all modes with wave vectors above the initial Nyquist
  frequency, the constraints are not violated by adding small scale power.}.  

This procedures allows for instance to
generate initial conditions which are consistant with, say a given
merging event, or the structure of the Local Group, the large scale
structures derived from large surveys such as the SDSS
(\citealt{2007arXiv0707.3413A}), the 2dF
(\citealt{2001MNRAS.327.1297P}), 2Mass
(\citealt{2003AJ....125..525J}), the Local Group
(\citealt{2005ApJ...635L.113M}), etc.

The ensemble of unconstrained gaussian random fields is defined by
all possible realizations of the field values $\delta({\bf x})$,
or their Fourier amplitudes $\delta({\bf k})$, for a given power
spectrum. If we impose the constraints on the field, the statistical
ensemble narrows down to a subset of realizations, those that have
the constraints satisfied. In particular, for a discrete representation
of the field on a grid of size $N^{3}$, this means that not all $N^{3}$
values of $\delta({\bf x})$ are real degrees of freedom. Averaging
over the constrained ensemble $\langle\ldots\rangle_{c}$ makes both
the mean $\langle\delta({\bf x})\rangle_{c}$ and the variance dependent
on position ${\bf x}$.

We shall be dealing with linear constraints each of which, in general,
sets a linear functional relation $V_{a}[\delta({\bf x})]$ (we will
use latin letters to index the constraints where each constraint is
defined by a grid position ${\bf x}_a$ and a constraint operator
$Y_a$, see below) between field values to
a given value, $V_{a}[\delta({\bf x})]=\tilde{V}_{a}$.

\begin{table*}
\begin{centering}
\begin{tabular}{lll}
Fourier space  & Configuration space  & Physical meaning \tabularnewline
\hline &  & \tabularnewline
$Y({\bf k})=1$  & $V[\delta]=\delta$  & density value \tabularnewline
$Y({\bf k})=-\imath{\bf k}_{i}/k^{2}$  & $V[\delta]=-\nabla_{i}\Delta^{-1}\delta$  & linear displacement \tabularnewline
$Y({\bf k})={\bf k}_{i}{\bf k}_{j}/k^{2}-\frac{1}{3}\delta_{ij}$  & $V[\delta]=(\nabla_{i}\nabla_{j}\Delta^{-1}-\frac{1}{3}\delta_{ij})\delta$ & flow shear \tabularnewline
$Y({\bf k})=\imath{\bf k}_{i}$ & $V[\delta]=\nabla_{i}\delta$  & density gradient \tabularnewline
$Y({\bf k})=-{\bf k}_{i}{\bf k}_{j}$ & $V[\delta]=\nabla_{i}\nabla_{j}\delta$ & density curvature \tabularnewline
\end{tabular}
\par\end{centering}
\caption{Different type of point -like constraints\label{tab:point-like constraints}}

\end{table*}

Alternatively, we can take a point of view that such a restricted
ensemble defines a new (constrained) random field $\delta_{c}({\bf x})$
which is still gaussian (due to the linearity of imposed relations)
but is statistically inhomogeneous. Under these conditions, the well
known way to construct a constrained field from unconstrained realizations
is \begin{equation}
\delta_{c}({\bf x})=\delta({\bf x})+\sum_{ab}\langle\delta({\bf x})V_{a}[\delta]\rangle\langle V_{a}[\delta]V_{b}[\delta]\rangle^{-1}\left(\tilde{V}_{b}-V_{b}[\delta]\right)\,.\end{equation}
 (see e.g. \citealt{1986ApJ...304...15B,hoffman}). In this expression,
the random quantities on the right-hand side are unconstrained $\delta(x)$.
Here $\tilde{V}_{b}$ is a numerical value of the constraint $b$,
while $\langle\delta({\bf x})V_{a}[\delta]\rangle$ is the covariance
between the field and a constraint, and $C_{ab}=\langle V_{a}[\delta]V_{b}[\delta]\rangle$
is the convariance matrix between the constraint functionals.

\begin{figure}[tbh]
\begin{centering}
\includegraphics[width=0.9\columnwidth]{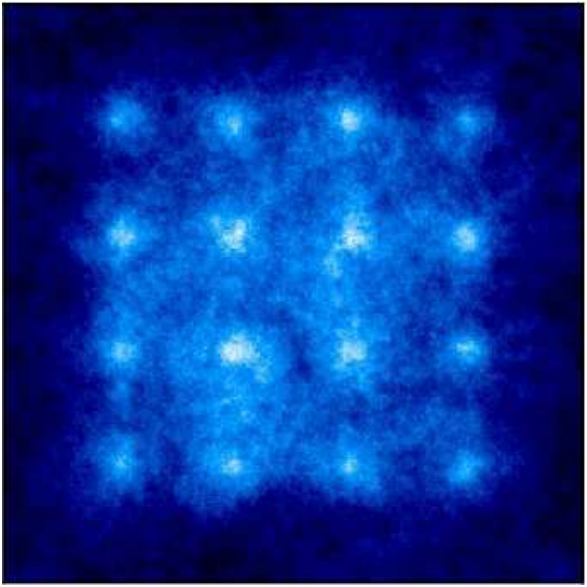}
\par\end{centering}

\caption{Example of contrained realization generated using \noun{constrfield}
and extended at a $1024^{3}$resolution using \texttt{mpgrafic}. Here
a regular grid of $4\times4\times4$ density peaks is imposed within a $\Lambda$CDM
cosmological simulation in a box of length $100h^{-1}$Mpc. Here the
constraints are of the density type, so that $Y({\bf k})=1$ for each
constraint; in such a case, $V_a(\delta)=\delta({\bf x}_a$),
 where $delta(
{x}_a)$ is the chosen constraint
at position $x_a$.\label{fig:Example-of-contrained}}

\end{figure}

This recipe reproduces the mean (note, the averaging is taken over
all unconstrained realizations) \begin{equation}
\langle\delta_{c}({\bf x})\rangle=\langle\delta({\bf x})V_{a}\rangle\langle V_{a}V_{b}\rangle^{-1}\tilde{V}_{b}\,,\end{equation}
 and the correlation function \begin{eqnarray}
\langle\delta_{c}({\bf x})\delta_{c}({\bf x^{\prime}})\rangle & = & \xi({\bf x,x^{\prime}})-\langle\delta({\bf x})V_{a}\rangle\langle V_{a}V_{b}\rangle^{-1}\langle\delta({\bf x^{\prime}})V_{b}\rangle\nonumber \\
 &  & +\langle\delta_{c}({\bf x})\rangle\langle\delta_{c}({\bf
  x^{\prime}})\rangle\,,
\label{eq:constr-xi}\end{eqnarray}
 which define all statistical properties for the gaussian case.

In cosmology the primary interest is to define constraints that describe
the physical properties of a patch (see e.g. \citealt{bondmyers,1996ASPC...94...49V})
of the density field - density, density derivatives, shear flow, averaged
over the volume of the patch. Such constraints can be represented
in Fourier space as \begin{equation}
V_{a}[\delta]=\int d^{3}k\delta({\bf k})Y_{a}({\bf
  k})W(kR_{p})e^{-{\bf ikx}_{a}}\,,
\label{eq:filtered-constraint}\end{equation}
 where ${\bf x}_{a}$ is the position of the patch, $W(kR_{p})$ is
a averaging filter over the patch size $R_{p}$, and $Y_{a}({\bf k})$
is the Fourier representation of the operator that specifies the constraint
functional. In particular, we use the constraint types given in Table
\ref{tab:point-like constraints}. Using constrained field formalism
\citet{1996Natur.380..603B} have demonstrated that the observed filamentary
Cosmic Web of matter distribution in the Universe can be understood
as dynamical enhancement of the geometrical properties of intial density
field. The web is largely defined by the position and primordial tidal
fields of rare events in the medium, such as precursors of large galaxies
at high redshifts or clusters of galaxies at present time, with the
strongest filaments between nearby clusters whose tidal tensors are
nearly aligned.

The code \texttt{constrfield} implements the same cosmology as \texttt{mpgrafic}
(including baryon wiggles) and offers the possibility of whitening
\footnote{Here we understand by whitening the operation that
  transforms an unconstrained field into a white noise, i.e. a
  collection of independent indentically distributed random variables
  $\tilde\mathcal{N}(0,1)$. Note that the presence of constraints
  breaks the independence of the grid cells even after ``whitening''
  (see Equation~\ref{eq:constr-xi}).} the resulting field in order to
feed it to \texttt{mpgrafic} as a low frequency input. An example of
this procedure is illustrated in
Fig. \ref{fig:Example-of-contrained}.

\subsection{Splitting the ICs}

Starting a parallel computation requires the initial conditions to be
dispatched over all the computing processes. Two alternatives exist to
perform this operation. The first one involves having the initial
conditions to be read by a `master' process and the data to be
broadcasted to all the other processes, and then keep or reject the
broadcasted data according to the topology of the computation's domain
split.  While simple to implement, this option happens to be difficult
to use in practice since broadcasting over a large number of processes can
be technically problematic and time consuming.

We present an alternative option which involves having the processes
upload their own set of data only. Because it is wasteful for each
process to parse the whole set of initial conditions to get the
relevant subset of data, this option implies that initial conditions
are pre-split according to the domain decomposition strategy of the
integrator. This splitting is both a domain boundary assignment,
a procedure for distributing particles among the processes, and a per
process  file dump.

 It results in a faster procedure, since each process
reads its own set of data, instead of having one process reading
the full initial conditions. A possible drawback lies in the fact
that a splitting is defined a priori; changing the number of processes
dedicated to the computation therefore requires the production of
a whole new set of split initial conditions. However splitting can
be performed on a single process, prior to any parallel computation,
and exhibits a negligible cost in terms of CPU-time. Such a procedure
can thus be applied an arbitrary number of times at almost no expense.

\begin{figure}[tbh]
\begin{centering}
\includegraphics[width=0.9\columnwidth]{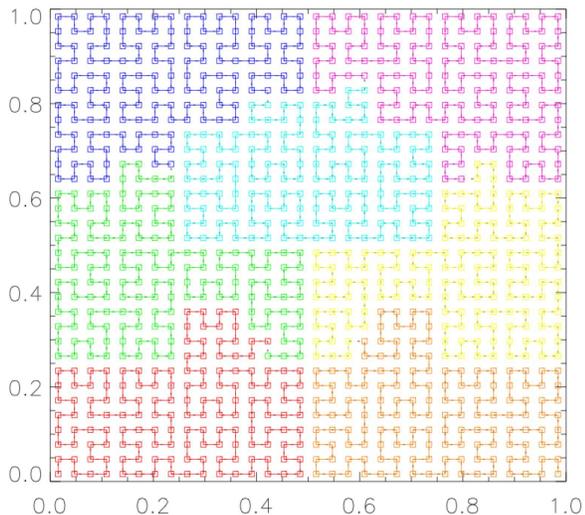}
\par\end{centering}

\caption{Example of a ICs splitting following a Peano-Hilbert domain decomposition
in two dimensions.}

\label{fig:peano2D}
\end{figure}

Horizon simulations were started from split initial conditions, where
each process uploaded its own set of data. The splitting scheme followed
the domain decomposition's strategy of the cosmological calculations
performed with RAMSES. It relies on a 3D Peano-Hilbert space filling
curve (\citealt{salmon1997poc,macneice2000ppa}) which provides a
complete mapping between the 3D position of a grid point and a 1D
coordinate on this curve. A two-dimensionnal example of
such a curve is shown in figure \ref{fig:peano2D}. By using this
piecewise linear representation of the computation domain, each process
is being given a continuous portion of this curve and load balancing
is achieved by `sliding' the limits of the local data sets along the
space-filling curve. In particular, the initial conditions are by
construction well-balanced, therefore the splitting among all the
processes involves a set of even subdivisions of the Peano-Hilbert
curve. A $(i,j,k)$ grid point maps to a single \textit{key} $q$.
A set of successive $(q_{1},q_{2},...,q_{n})$ corresponds to a single
process $p$.

\begin{figure}[tbh]
\begin{centering}
\includegraphics[width=0.9\columnwidth]{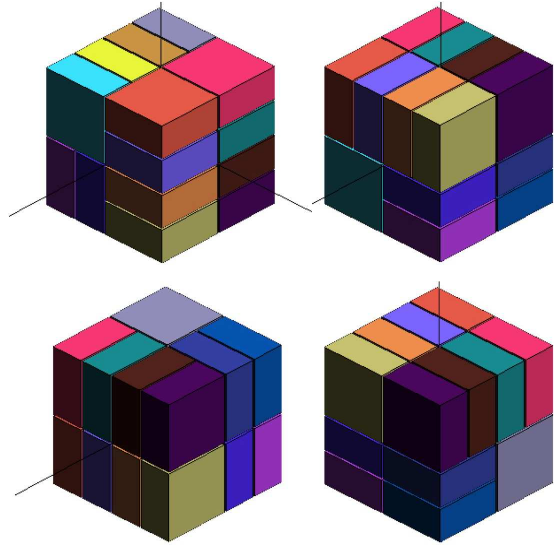}
\par\end{centering}

\caption{Example of a ICs splitting in 16 subvolumes following a Peano-Hilbert
domain decomposition as seen from four different directions.}

\label{fig:peano3D}
\end{figure}

Note that all sub-domains are simply connected, i.e. within there
are no isolated sub-regions owned by a process in the middle of
another region owned by another process. For instance, in 3D, if
the curve is split in $2^{r}$ sections, each section fits in a 3D
rectangle of different sizes and orientations (see figure \ref{fig:peano3D}).
Moreover, if the curve is split in $8^{r}$ parts, all the sections
fit in a cube of the same volume. This type of domain decomposition
is known to be most efficient if we consider the ensemble of all the
refined grids configurations. It may be surpassed by other strategies
(e.g. layer splitting, angular splitting) on specific situations,
but Peano-Hilbert domain decomposition remains the best strategy \textit{on
average}.

Here we implemented a fast and simple algorithm which performs the
splitting of initial conditions according to the Peano-Hilbert domain
decomposition. It relies on a plane-by-plane investigation of the
data cube which limits the memory consumption. Let $n^{3}$ be the
number of cells in the initial conditions data cube, $F(i,j,k)$ the
value of the 3D field at grid indices $1\leq i,j,k\leq n$ and $f$
a 2D slice of $F$ at index $k$. We assume that the number of processes
nproc is a \textit{power of two} (even though this constraint can
be lifted as explained below); it ensures that each process domain
is a 3D rectangle. Consequently the extent of the sub-domain corresponding
to a process $p$ is given by two triplets $(i_{m},j_{m},k_{m})^{p}$
and $(i_{M},j_{M},k_{M})^{p}$ which correspond to the two extreme
corners of the rectangle. The algorithm described below runs on one
process only, and involves two distinct steps:

\begin{itemize}
\item do $k=1,n$ : loop over data planes 

\begin{itemize}
\item read $f=F(1:n,1:n,k)$ : the 2D data is uploaded. 
\item initialisations: 
\begin{itemize}
\item process$(1:\mathrm{nproc})$ =.false.,
\item  $(i_{m},j_{m},k_{m})^{p}=-1$, $(i_{M},j_{M},k_{M})^{p}=-1$,\,
  $\forall p \in [1:\mathrm{nproc}]$
\end{itemize}
 
\item do $j=1,n$ , $i=1,n$ 

\begin{itemize}
\item Peano-Curve mapping : (i,j,k) $\rightarrow$ q $\rightarrow$ p 
\item process p is found in the k plane: process(p)=.true. 
\item if process p found for the 1st time set the minimum position : if $(i_{m},j_{m},k_{m})^{p}=-1$
then $(i_{m},j_{m},k_{m})^{p}=(i,j,k)$ 
\item update the maximum position: if $(i,j,k)>(i_{M},j_{M},k_{M})^{p}$
then $(i_{M},j_{M},k_{M})^{p}=(i,j,k)$ 
\end{itemize}
\item do $p=1,\textrm{nproc}$ 

\begin{itemize}
\item if process(p)=.false. skip : only processes detected in the plane are taken
in account. 
\item write f($i_{m}^{p}:i_{M}^{p},j_{m}^{p}:j_{M}^{p}$) in the file of
the process p.
\end{itemize}
\end{itemize}
\end{itemize}

First, initial conditions fields are parsed plane by plane and for
each plane, the process map is achieved through the space-filling curve
mapping. Then for each process found in the current plane the subset
of data is written in the relevant files. Because of the compacity
of the sub-domains, a significant speedup can be obtained by parsing
the i and j indexes of the second loop with steps larger than 1~:
this procedure is safe as long as the step remains smaller than the
smallest extent of the sub-domains along one direction. 
We call this step the speedup step. Finally the
nproc$=2^{r}$ constraint can be lifted if a set of processes upload
the same set of data. Each process would load a small initial condition
file and would retain only its `sub-sub-domain' within a sub-domain.
For instance the \noun{Horizon} 4$\Pi$ simulation ran on $6144=3\times2048$
processes: the splitting was performed over 2048 sub-domains and
each sub-file was uploaded by three processes.
 Overall, this algorithm
can be quite effective and as an illustration, the $4096^{3}$ initial
conditions fields of this simulation were split in 15 minutes on a
single process of the CCRT computing center using a speedup step
of $256$. The code ran on Itanium2 processors
   (double-core, but only a single core has been used here) with a $1.6$
   GHz frequency.

\section{Application to large Horizon simulations}
\label{sec:Application-to-large}

Let us now illustrate on a couple of large scale simulation how\texttt{
mpgrafic }was used. We will consider in turn a hydrodynamical and
a dark matter only simulation.

\subsection{Horizon-MareNostrum}

The first major application of the \texttt{mpgrafic} code was the
generation of ICs for a simulation of a cosmological hydrodynamical
simulation of linear size $50h^{-1}$Mpc on a grid of size $1024^{3}$. 

For the Mare Nostrum simulation, we started for technical reasons
with external ICs (given as a set of particle velocities), as one
of the goals was to compare two n-body plus hydro codes (namely the
RAMSES and GADGET2 codes) on similar ICS. The particle velocities
were therefore read from external ICs and we performed the derivation
of the density field samples on the grid from the particle velocities
in Fourier space, using the JMFFT Fourier Transform package%
\footnote{\texttt{http://www.idris.fr/data/publications/JMFFT/}%
}. The $f$ factor involved in Eq. (\ref{eq:deff})  was computed using
routines provided in the original \texttt{Grafic1} package. Once the
velocity divergence and $f$are known, the initial density field is
easily recovered and ready to be used as an input for simulations
(especially for the hydrodynamical part of RAMSES), or as a source
for a specific set of initial phases.

\begin{figure}[tbh]
\begin{centering}
\includegraphics[width=0.9\columnwidth]{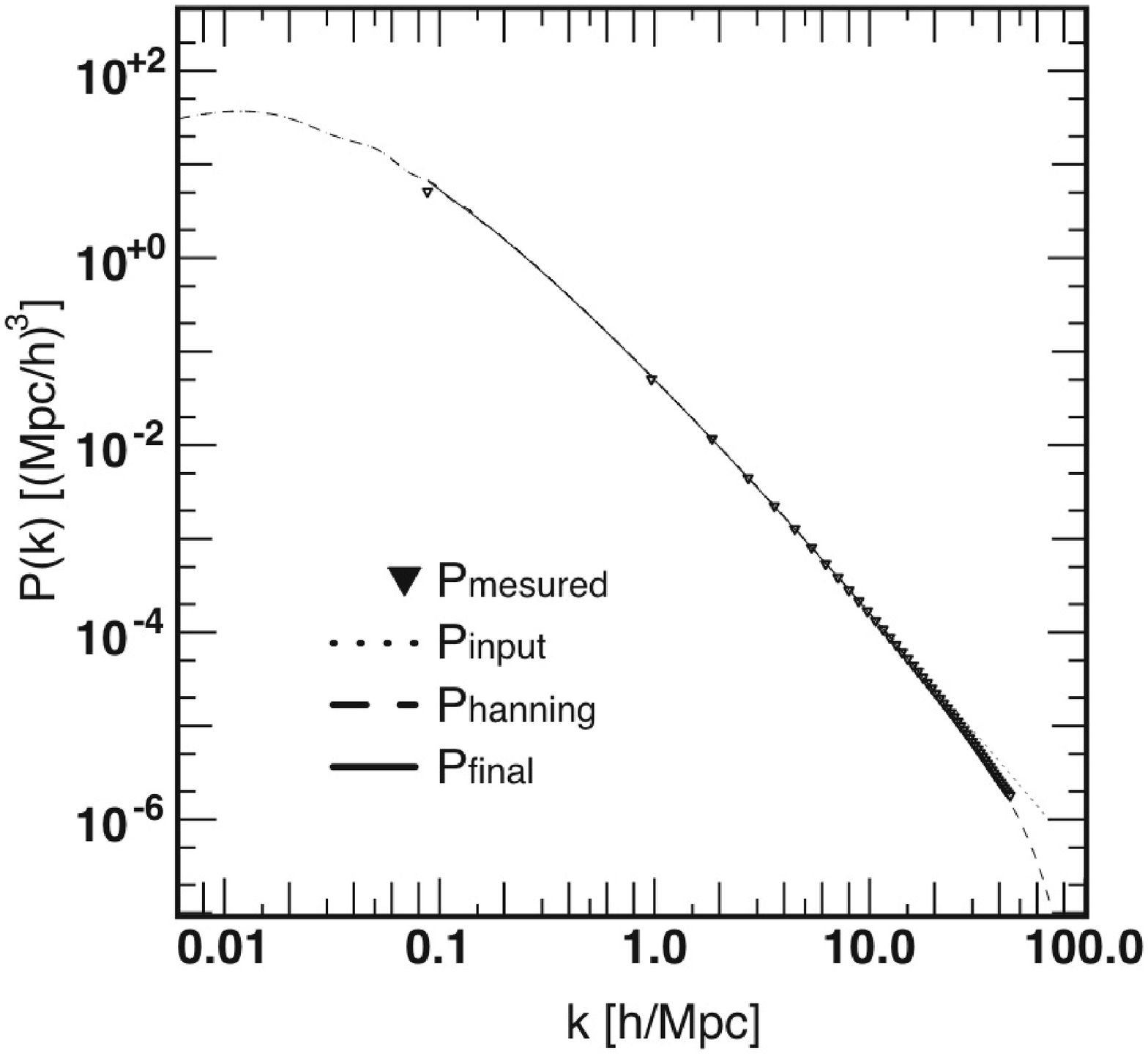} \includegraphics[width=0.83 \columnwidth]{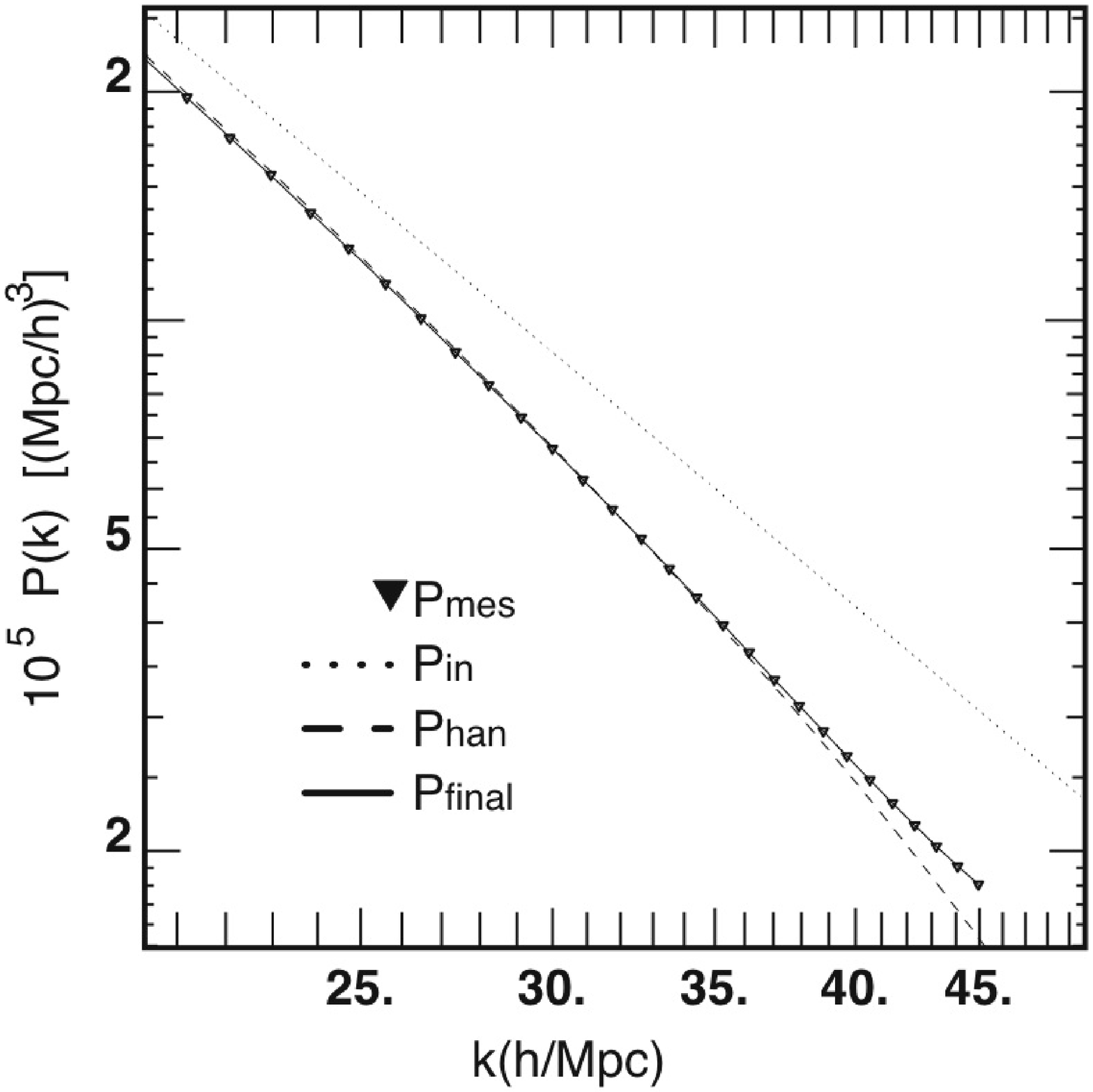} 
\par\end{centering}

\caption{The power spectrum $P(k)$ of the 'Mare Nostrum' initial conditions.
Symbols stand for the measured power spectrum, $P_{\mathrm{mes}}(k)$, while the dotted line
stands for the theoretical power spectrum $P_{\mathrm{in}}(k)$. The dashed
line, $P_{\mathrm{Han}}(k)$ , stand for the
theoretical power spectra plus a Hanning filter contribution. The
solid line, $P_{\mathrm{final}}(k)$ stand for our best fit of the power spectrum. The bottom 
panel represents a zoomed version of the top one where only the small
scales are shown.}

\label{f:pkfit} 
\end{figure}

\subsubsection{Power Spectrum Extraction from Mare Nostrum Initial Conditions}

\begin{figure*}[t]
\begin{centering}
\includegraphics[width=1.9\columnwidth]{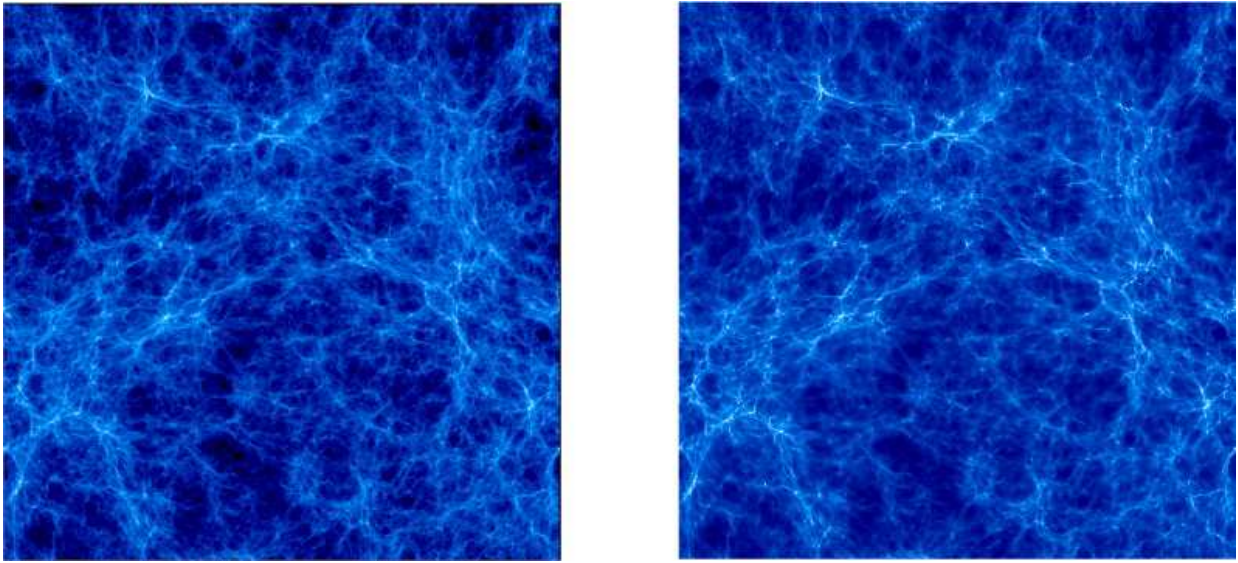}
\par\end{centering}
\caption{Comparison between the gas in the SPH \noun{MareNostrum} simulation
at redshift $z=5.7$ (\emph{left}) from which ICs the Horizon white
noise was extracted, and the Horizon\noun{-MareNostrum} simulation
(\emph{right}) for which the initial conditions were generated with
\texttt{mpgrafic}. This figure demonstrates that on large scales the
phases are indeed reproduced by the procedure. It also shows that
on these scales, the two code produce quite similar features.\label{fig:Comparison-between-the}}
\end{figure*}

The master equation (Eq. (\ref{e:master})) can only be inverted knowing
the convolution kernel, i.e. the power spectrum $P(k)$. In principle,
the knowledge of the cosmology and the included physics should be
sufficient to derive $P(k)$ prior to the deconvolution. Let us call
$P_{\mathrm{in}}(k)$, this theoretical power spectrum constrained
only by physics. In practice, this theoretical power spectrum differs
from the effective power spectrum used to draw the (external, particle
based) ICs. Our goal in this section is to define a power spectrum
$P_{\mathrm{final}}(k)$ that should accurately represent the ensemble
averaged power spectrum of the external ICs (so that $\langle
P_{\mathrm{mes}}(k)\rangle \approx P_{\mathrm{final}}(k))$), based on
the available theoretical and measured power spectra. 
$P_{\mathrm{final}}(k)$ will then be used in the deconvolution. 


Using an inaccurate spectrum to deconvolve Eq. (\ref{e:master}) would
lead to a 'colored' noise for the initial phases, i.e. with spurious
characteristic length scales.

Let us illustrate the discrepancy between the theoretical and the
measured $P(k)$ by describing the power spectrum of the Mare Nostrum
ICs. The measured $P(k)$ is shown in figure \ref{f:pkfit} as triangles.
The theoretical $P(k)=P_{\mathrm{in}}(k)$ (i.e. the one used to generate
this set of ICs) is also shown as a dotted line and unsurprisingly,
the two curves disagree.

At low $k$, the finite volume of the simulation implies that the
empirical power spectrum of large scale modes has large sampling variance
and thus departs from the theoretical curve. The high $k$ discrepancy
is of different nature: clearly the sampling variance is negligible,
but now the discreetness of the grid and anisotropy of very high $k$
mode representation play role. $P(k)$ departs significantly from
$P_{\mathrm{in}}(k)$ as easily seen when zooming on the large k regions,
where $P(k)$ lacks power compared to the expected behavior. Therefore,
$P_{\mathrm{in}}(k)$ cannot be used without corrections to whiten the
external IC set. 
In the following, we  rely on the fact that Gaussian initial 
conditions are statistically characterized by power spectrum only.

The exact set of corrections depends on how the field have been
generated.  For Mare Nostrum IC's we must first include a Hanning
filter defined in the Fourier space by \begin{equation}
W_{H}(k)=\cos(\frac{\pi k}{2k_{N}}),\end{equation} Here the Nyquist
frequency is given by $k_{N}=2\pi/L\times N/2$, where $L=50$ Mpc/h is
the size of the box of the Mare Nostrum ICs and N stands for the
(linear) number of grid elements. Such a filtering is frequently
encountered when dealing with initial conditions: because Fourier
modes are sampled on a cartesian grid, the two conditions $k<\pi
N\sqrt{3}/L$ and $(k_{x},k_{y},k_{z})<\pi N/L$ imply that anisotropies
arise on the smallest scales along the diagonals. The Hanning filter
damps high frequency modes, and reduces the small scales contributions
and consequently the anisotropies. In Fig. \ref{f:pkfit}, we display the
$P_{\mathrm{Han}}$ curve as a dashed line, where: \begin{equation}
P_{\mathrm{Han}}=W_{H}(k)^{2}P_{\mathrm{in}}(k).\end{equation}
Clearly, $P_{\mathrm{Han}}(k)$ reproduces well the measured power
spectrum (except at high frequencies, see below), with $N=2048$, which
corresponds to the original resolution of the external ICs, prior to
some (external) degradation procedure. This modification of the
spectrum corresponds to the most favorable case where an analytic
expression is known or can be found for the filtering applied on the
data.

Secondly, Fig. \ref{f:pkfit} shows that $P_{\mathrm{Han}}(k)$ still
lack some power for $k>40$ h/Mpc. This feature corresponds to the
external degradation procedure: one particle out of eight was provided,
out of the original $2048^{3}$ particles, resulting in power aliasing
at high frequency. This part of the power spectrum was fitted by a
smoothed version of the measured power spectrum. We call $P_{\mathrm{final}}(k)$
this final power spectrum that includes the effect of the Hanning
Filter and corrects the high frequencies effects due to the degradation:
\begin{eqnarray}
P_{\mathrm{final}}(k)= & P_{\mathrm{Han}}(k) & \mathrm{k}\le40\mathrm{h/Mpc},\\
 & S[P_{\mathrm mes}(k)] & \mathrm{k}>40\mathrm{h/Mpc},\end{eqnarray}
 where $S[X]$ stands for a smoothing operator. By using a smoothed
version of the spectrum, we avoid overfitting of the fluctuations
in the spectrum, which would artificially reduce the variance at these
scales.

To conclude, $P_{\mathrm{final}}(k)$ combines both an analytic
expression of the filtered theoretical spectrum at low frequencies and
a numerical evaluation at high frequencies, based on the measured
power spectrum. We emphasize that these choices are by no means unique
but were found to provide phases with the proper spectrum.

\subsubsection{Whitening}

\begin{figure}[tbh]
\begin{centering}
\includegraphics[width=0.9\columnwidth]{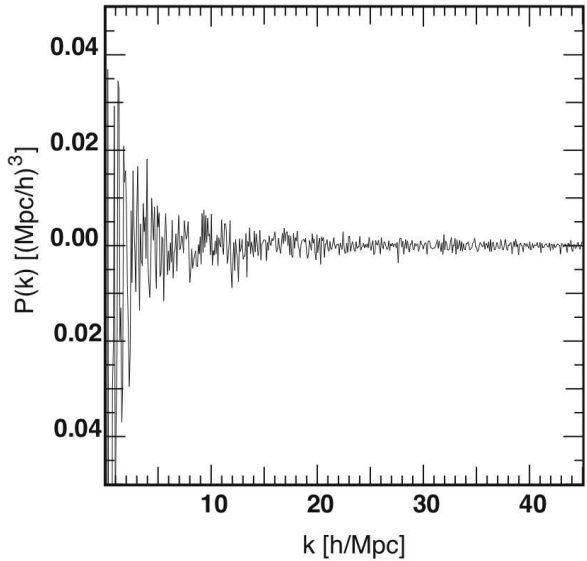} 
\par\end{centering}

\caption{The power spectrum of the phases contained in the 'Mare Nostrum' initial
conditions, with the expectation value of the power of a white noise
of unit variance substracted . As expected from a white noise's realisation,
the spectrum fluctuates around an overall flat line. }

\label{f:wnpk} 
\end{figure}

The {}``whitening'' operation (i.e. getting the white noise $n_{1}({\bf x})$
from $\delta({\bf x})$ and $P_{\mathrm{final}}(k)$) is then performed
by deconvolution as in Eq. (\ref{e:master}). For the Horizon Mare
Nostrum ICs, the power spectrum of the resulting white noise is shown
in Fig. \ref{f:wnpk}. The whitened ICs are now ready to be processed
into new initial conditions: the whitened ICs serve as low frequency
constraints when generating the refined ICs using \texttt{mgrafic}

The code \texttt{mpgrafic} was then used to generate the ICs of a
$\Lambda$CDM hydrodynamical {}``full physics'' simulation (with
star formation and metals) with a boxsize of $50h^{-1}$Mpc on a grid
of $1024^{3}$ using the \noun{Horizon} reference white noise. The
comparison between the input phases and the reproduced phases with
\texttt{mpgrafic} is illustrated in Figure \ref{fig:Comparison-between-the}
which shows both simulations at redshift 5.7.

\subsection{\noun{Horizon 4$\Pi$}}

\begin{figure}[tbh]
\begin{centering}
\includegraphics[clip,width=0.9\columnwidth]{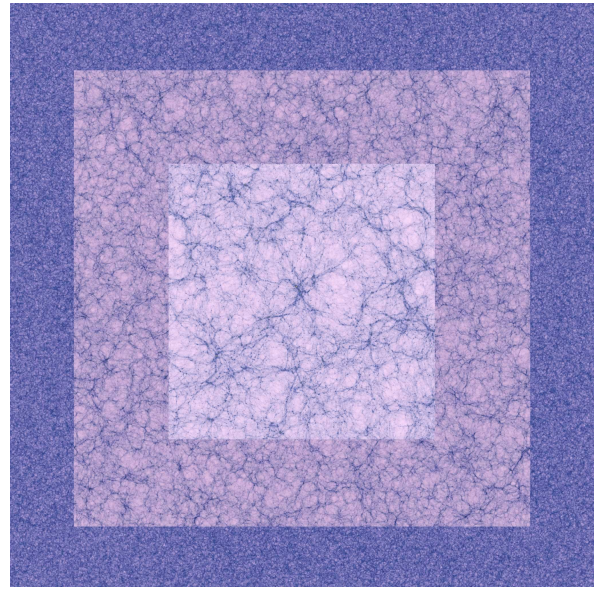}
\par\end{centering}
\caption{A multi resolution view of HORIZON 4$\Pi$. The outer region corresponds
to a view of the universe on scales of $16h^{-1}$Gpc: it is generated
by unfolding the simulation while cuting a slice obliquely through
the cube in order to preserve the continuity of the field thanks to
the periodicity. The intermediate region corresponds to a slice of
$2h^{-1}$Gpc, while the inner region is at the original resolution
the initial conditions. \noun{RAMSES} has refined 6 times over the
course of the run from that resolution.}

\end{figure}
\begin{figure}[tbh]
\begin{centering}
\includegraphics[width=0.9\columnwidth]{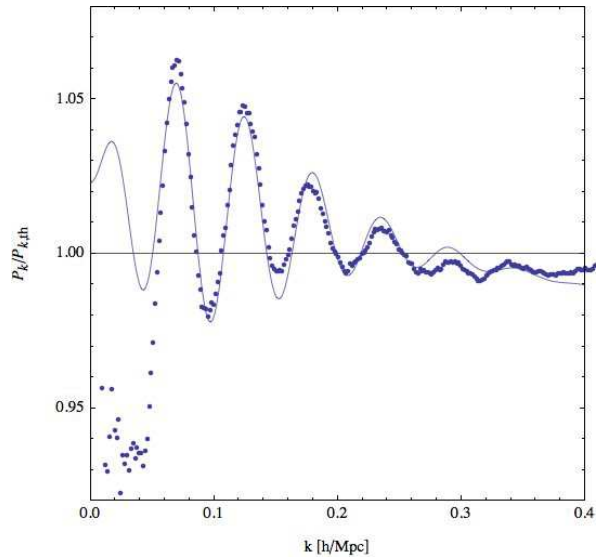}
\end{centering}
\caption{The measured baryon wiggles at $z=0$  together with the corresponding
fit (scaling like $\exp(-[k/0.1]^{1.4})\sin(2\pi k/k_{A})$ plus some
linear drift in $k$), which finds that $2\pi/k_{A}=113\: h^{-1}$Mpc.
\label{fig:BAO}}
\end{figure}
\begin{figure}[H]
\begin{centering}
\includegraphics[clip,width=0.9\columnwidth]{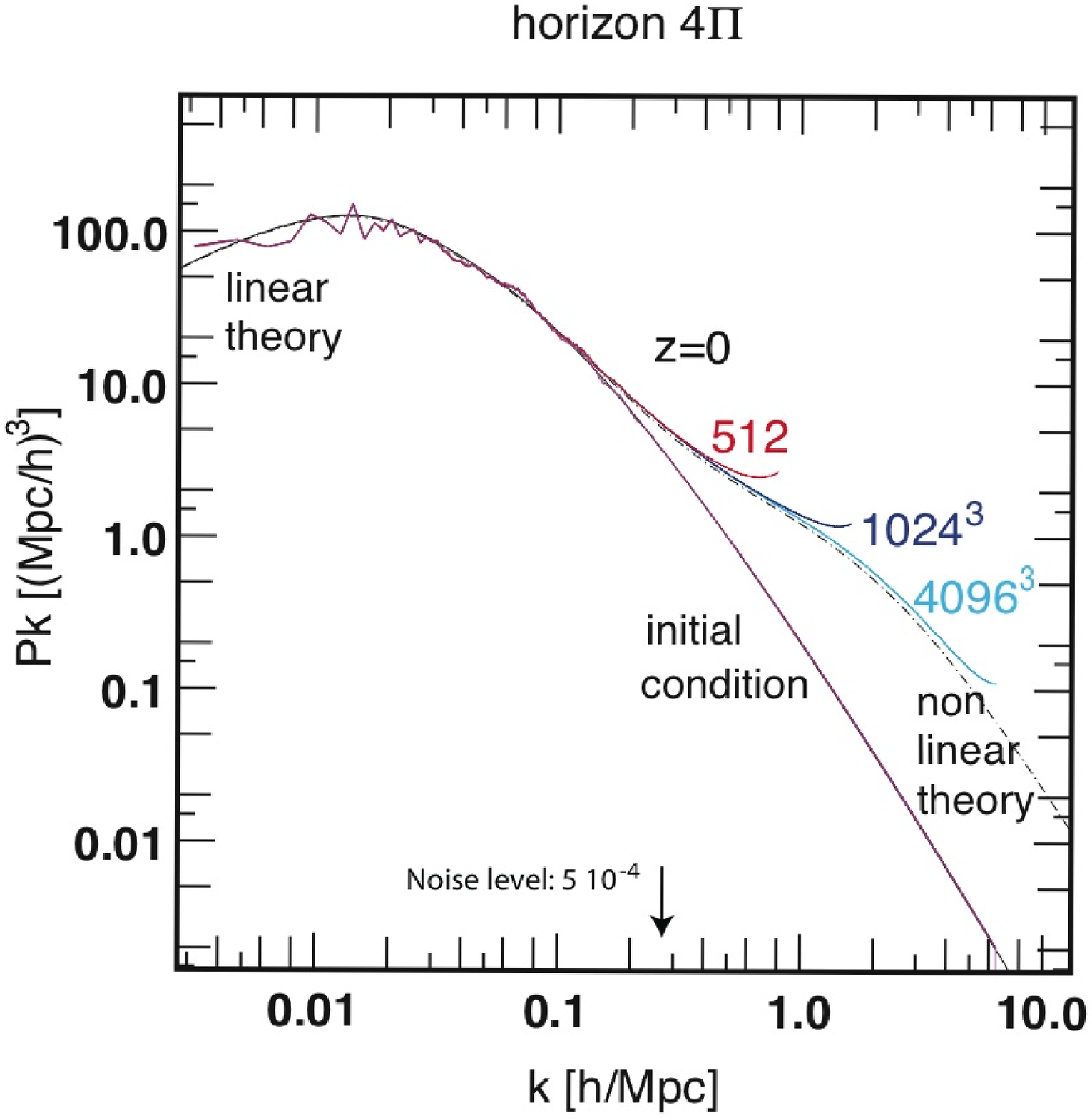}
\end{centering}
\caption{A Measure of the power spectrum with \texttt{powergrid} of the \noun{Horizon-$4\Pi$}
in the \texttt{mpgrafic} generated initial condition (\emph{bottom
curve}) and at redshift zero for different samplings (\emph{top curves
}for resp. $512^{3},1024^{3}$and $4096^{3}$ as labeled), with the
modes $k$ in units of $h^{-1}$Mpc. Here the density is resampled
on the grids using the nearest grid point kernel (NGP) whose bias
is corrected. However, the shot noise bias was not corrected in this
figure, and no attempt was made to correct the power aliased by the
resampling procedure. Both measurements are compared to resp. the
linear theory and the theoretical predictions of \citet{smith}. Note
that the agreement between the linear theory and the generated initial
conditions is excellent for all measured scales. \label{fig:4pipowerspec}}
 \end{figure}

As mentionned earlier, the code \texttt{mpgrafic} was also used to
generate the ICs of \noun{Horizon}-$4\Pi$ a $\Lambda$CDM dark matter
only simulation based on cosmological parameters inferred by the WMAP
three-years results, with a boxsize of $2h^{-1}$Gpc on a grid of
size $4096^{3}$. The purpose of this simulation is to investigate
full sky weak lensing and baryonic accoustic oscillations. The 70
billions particles were evolved using the Particle Mesh scheme of
the \emph{\noun{RAMSES}} code on an adaptively refined grid (AMR)
with about 140 billions cells. Each of the 70 billions cells of the
base grid was recursively refined up to 6 additional levels of refinement,
reaching a formal resolution of 262144 cells in each direction (roughly
7 kpc/h comoving).

The corresponding power spectrum was measured using \texttt{powergrid},
and is shown in Figure \ref{fig:4pipowerspec}. Since the simulation
snapshots involves a collection of particles, we had to resample them
on a grid using a convolution kernel before estimating the power spectrum.
The resulting (analytical) bias in the power spectrum was corrected;
however, this resampling procedure leads to some power aliasing close
to the Nyquist frequency that cannot be corrected without additional
information. Finally, note that the AMR structure of the \noun{RAMSES}
code leaves the opportunity of measuring the power spectrum at frequencies
beyond the Nyquist frequency of the $4096^{3}$ grid. Such measurements
are outside the scope of this paper, and the power spectrum tool \texttt{powergrid}
should be viewed primarily as a diagnosis tool. Here, as a check of
both the initial condition generation algorithm, including the implementation
of the baryon wiggles, a novelty of this implementation, figure \ref{fig:BAO}
displays the measured baryon wiggles in the $z=0$ snapshot, together
with the corresponding fit.

\section{Conclusion}

A series of tools to construct and validate initial conditions for
large ($n\ge1024^{3}$) cosmological simulations in parallel were
presented and illustrated. These tools involve ICs generation with
optional constraints, low-pass filtering and resampling, power spectrum
estimation, estimation of matter density on a grid from a set of particle
positions and Peano-Hilbert domain decomposition. As illustrated in
section \ref{sec:Application-to-large}, they allowed us to produce
very large cosmological simulations. From these high-resolution ICs,
one can then create at will, zoom-like initial conditions and constrained
ICs with the help of the resampling tool. Let us emphasize that \texttt{mprafic}
provides an alternative route to initial condition generation such
as \texttt{Grafic}2. It is more versatile as it does not impose any
relationship between resolution and boundaries for the refined sub
volumes. The remaining limitation is the total amount of memory available
on distributed architectures. A logical extension of this work will
be to generate initial conditions corresponding to the local group.
Note finally that with simple amendments, the above mentionned code
could be used in the context of vector field generation (magnetic
field with a given helicity), or turbulence.

\acknowledgements

We warmly thank the Barcelona Supercomputing Center and the CCRT staff
for their help in producing the Horizon-MareNostrum and the Horizon-4$\Pi$
simulations. We also thank the referee for his careful reading of the
manuscript, G. Yepes for his help with the MareNostrum
initial conditions, Stephane Colombi, Karim Benabed, Julien Devriendt,
Thierry Sousbie for advices, and D.~Munro for freely distributing
his \texttt{Yorick} programming language and opengl interface (available
at \url{http://yorick.sourceforge.net}). This work was carried
within the framework of the \noun{horizon} project:
\url{http://www.projet-horizon.fr}. All codes described in this paper
are available at \url{ftp://ftp.iap.fr/pub/from\_users/prunet/}.

\appendix

\section{MPgrafics parameter file\label{sec:Running-constfield}}

\begin{table*}
\begin{centering}
\begin{tabular}{||lc|l||}
\hline 
Parameters  &  & Description \tabularnewline
\hline 
$4$  &  & Power spectrum parametrization: $4$ is Eisenstein \& Hu \tabularnewline
$0.24,0.76,73.0$  &  & $\Omega_{m},\Omega_{\Lambda},H_{0}$\tabularnewline
$0.042$  &  & $\Omega_{b}$ \tabularnewline
$0.96$  &  & $n_{S}$ \tabularnewline
$-0.92$  &  & Normalization: $-\sigma_{8}$ if negative, $Q_{rms}$ if positive \tabularnewline
$0.01,100.0$  &  & $k_{min}$ and $k_{max}$ in $h^{-1}Mpc$ for analytical PS output \tabularnewline
$-50.0$  &  & Box length in $h^{-1}Mpc$ if negative, mesh length in $Mpc$ if positive\tabularnewline
$1$  &  & \texttt{Grafic1} mode: no choice here\tabularnewline
$0$  &  & \texttt{Grafic1} mode: no choice here\tabularnewline
 &  & \tabularnewline
 &  & \tabularnewline
 &  & \tabularnewline
 &  & \tabularnewline
$1$  &  & $1$: Generate noise file and save it / $2$: Read from noise file\tabularnewline
$1234$  &  & Initial seed (useful if $1$ is set above)\tabularnewline
white-$256$.dat  &  & Noise file name \tabularnewline
$1$  &  & $1$: Constraint of large scale phases with small noise file / $2$:
No padding\tabularnewline
white-$128$.dat  &  & Small (constraint) noise file name\tabularnewline
\hline
\end{tabular}
\par\end{centering}
\caption{ \texttt{mpgrafic} parameter file example.\label{tab:paramfile}}

\end{table*}


\begin{thebibliography}{23}
\expandafter\ifx\csname natexlab\endcsname\relax\def\natexlab#1{#1}\fi
\expandafter\ifx\csname url\endcsname\relax
  \def\url#1{{\tt #1}}\fi

\bibitem[{Adelman-McCarthy} and {for the SDSS
  Collaboration}(2007)]{2007arXiv0707.3413A}
J.~K. {Adelman-McCarthy} and {for the SDSS Collaboration}.
\newblock {The Sixth Data Release of the Sloan Digital Sky Survey}.
\newblock {\em ArXiv e-prints}, 707, July 2007.

\bibitem[{Bardeen} et~al.(1986){Bardeen}, {Bond}, {Kaiser}, and
  {Szalay}]{1986ApJ...304...15B}
J.~M. {Bardeen}, J.~R. {Bond}, N.~{Kaiser}, and A.~S. {Szalay}.
\newblock The statistics of peaks of gaussian random fields.
\newblock {\em \apj}, 304:\penalty0 15--61, May 1986.

\bibitem[{Bertschinger}(2001)]{grafic2}
E.~{Bertschinger}.
\newblock {Multiscale Gaussian Random Fields and Their Application to
  Cosmological Simulations}.
\newblock {\em \apjs}, 137:\penalty0 1--20, November 2001.

\bibitem[{Bond} et~al.(1996){Bond}, {Kofman}, and
  {Pogosyan}]{1996Natur.380..603B}
J.~R. {Bond}, L.~{Kofman}, and D.~{Pogosyan}.
\newblock {How filaments of galaxies are woven into the cosmic web}.
\newblock {\em \nat}, 380:\penalty0 603--+, April 1996.

\bibitem[{Bond} and {Myers}(1996)]{bondmyers}
J.~R. {Bond} and S.~T. {Myers}.
\newblock {The Peak-Patch Picture of Cosmic Catalogs. I. Algorithms}.
\newblock {\em \apjs}, 103:\penalty0 1--+, March 1996.

\bibitem[{Cen} and {Ostriker}(2000)]{2000ApJ...538...83C}
R.~{Cen} and J.~P. {Ostriker}.
\newblock {Physical Bias of Galaxies from Large-Scale Hydrodynamic
  Simulations}.
\newblock {\em \apj}, 538:\penalty0 83--91, July 2000.

\bibitem[{Eisenstein} and {Hu}(1998)]{eisenstein}
D.~J. {Eisenstein} and W.~{Hu}.
\newblock {Baryonic Features in the Matter Transfer Function}.
\newblock {\em \apj}, 496:\penalty0 605--+, March 1998.

\bibitem[{Frenk} et~al.(2000){Frenk}, {Colberg}, {Couchman}, {Efstathiou},
  {Evrard}, {Jenkins}, {MacFarland}, {Moore}, {Peacock}, {Pearce}, {Thomas},
  {White}, and {Yoshida}]{2000astro.ph..7362F}
C.~S. {Frenk}, J.~M. {Colberg}, H.~M.~P. {Couchman}, G.~{Efstathiou}, A.~E.
  {Evrard}, A.~{Jenkins}, T.~J. {MacFarland}, B.~{Moore}, J.~A. {Peacock},
  F.~R. {Pearce}, P.~A. {Thomas}, S.~D.~M. {White}, and N.~{Yoshida}.
\newblock {Public Release of N-body simulation and related data by the Virgo
  consortium}.
\newblock {\em ArXiv Astrophysics e-prints}, July 2000.

\bibitem[{Gottl{\"o}ber} and {Yepes}(2007)]{2007ApJ...664..117G}
S.~{Gottl{\"o}ber} and G.~{Yepes}.
\newblock {Shape, Spin, and Baryon Fraction of Clusters in the MareNostrum
  Universe}.
\newblock {\em \apj}, 664:\penalty0 117--122, July 2007.

\bibitem[{Hoffman} and {Ribak}(1991)]{hoffman}
Y.~{Hoffman} and E.~{Ribak}.
\newblock {Constrained realizations of Gaussian fields - A simple algorithm}.
\newblock {\em \apjl}, 380:\penalty0 L5--L8, October 1991.

\bibitem[{Jarrett} et~al.(2003){Jarrett}, {Chester}, {Cutri}, {Schneider}, and
  {Huchra}]{2003AJ....125..525J}
T.~H. {Jarrett}, T.~{Chester}, R.~{Cutri}, S.~E. {Schneider}, and J.~P.
  {Huchra}.
\newblock {The 2MASS Large Galaxy Atlas}.
\newblock {\em \aj}, 125:\penalty0 525--554, February 2003.

\bibitem[MacNeice et~al.(2000)MacNeice, Olson, Mobarry, de~Fainchtein, and
  Packer]{macneice2000ppa}
P.~MacNeice, K.M. Olson, C.~Mobarry, R.~de~Fainchtein, and C.~Packer.
\newblock {PARAMESH: A parallel adaptive mesh refinement community toolkit}.
\newblock {\em Computer Physics Communications}, 126\penalty0 (3):\penalty0
  330--354, 2000.

\bibitem[{Mohayaee} and {Tully}(2005)]{2005ApJ...635L.113M}
R.~{Mohayaee} and R.~B. {Tully}.
\newblock {The Cosmological Mean Density and Its Local Variations Probed by
  Peculiar Velocities}.
\newblock {\em \apjl}, 635:\penalty0 L113--L116, December 2005.

\bibitem[{Ocvirk}, {Pichon} and {Teyssier}(2008)]{Ocvirk2008}
P.~{Ocvirk}, C.~{Pichon} and R.~{Teyssier}.
\newblock {Bimodal gas accretion in the Mare Nostrum galaxy formation simulation}.
\newblock submitted to {\em \mnras}

\bibitem[{Pen}(1997)]{pen}
U.-L. {Pen}.
\newblock {Generating Cosmological Gaussian Random Fields}.
\newblock {\em \apjl}, 490:\penalty0 L127+, December 1997.

\bibitem[{Percival} et~al.(2001){Percival}, {Baugh}, {Bland-Hawthorn},
  {Bridges}, {Cannon}, {Cole}, {Colless}, {Collins}, {Couch}, {Dalton}, {De
  Propris}, {Driver}, {Efstathiou}, {Ellis}, {Frenk}, {Glazebrook}, {Jackson},
  {Lahav}, {Lewis}, {Lumsden}, {Maddox}, {Moody}, {Norberg}, {Peacock},
  {Peterson}, {Sutherland}, and {Taylor}]{2001MNRAS.327.1297P}
W.~J. {Percival}, C.~M. {Baugh}, J.~{Bland-Hawthorn}, T.~{Bridges},
  R.~{Cannon}, S.~{Cole}, M.~{Colless}, C.~{Collins}, W.~{Couch}, G.~{Dalton},
  R.~{De Propris}, S.~P. {Driver}, G.~{Efstathiou}, R.~S. {Ellis}, C.~S.
  {Frenk}, K.~{Glazebrook}, C.~{Jackson}, O.~{Lahav}, I.~{Lewis}, S.~{Lumsden},
  S.~{Maddox}, S.~{Moody}, P.~{Norberg}, J.~A. {Peacock}, B.~A. {Peterson},
  W.~{Sutherland}, and K.~{Taylor}.
\newblock {The 2dF Galaxy Redshift Survey: the power spectrum and the matter
  content of the Universe}.
\newblock {\em \mnras}, 327:\penalty0 1297--1306, November 2001.

\bibitem[Salmon and Warren(1997)]{salmon1997poc}
J.~Salmon and MS~Warren.
\newblock {Parallel, out-of-core methods for N-body simulation}.
\newblock {\em Conference: 8. SIAM conference on parallel processing for
  scientific computing, Minneapolis, MN (United States), 14-17 Mar 1997}, 1997.

\bibitem[{Shirokov}(2007)]{2007arXiv0711.4655S}
A.~{Shirokov}.
\newblock {GRAvitational COSmology (GRACOS) code release announcement, for
  version 1.0.1a9}.
\newblock {\em ArXiv e-prints}, 711, November 2007.

\bibitem[{Smith} et~al.(2003){Smith}, {Peacock}, {Jenkins}, {White}, {Frenk},
  {Pearce}, {Thomas}, {Efstathiou}, and {Couchman}]{smith}
R.~E. {Smith}, J.~A. {Peacock}, A.~{Jenkins}, S.~D.~M. {White}, C.~S. {Frenk},
  F.~R. {Pearce}, P.~A. {Thomas}, G.~{Efstathiou}, and H.~M.~P. {Couchman}.
\newblock {Stable clustering, the halo model and non-linear cosmological power
  spectra}.
\newblock {\em \mnras}, 341:\penalty0 1311--1332, June 2003.

\bibitem[{Springel} et~al.(2005){Springel}, {White}, {Jenkins}, {Frenk},
  {Yoshida}, {Gao}, {Navarro}, {Thacker}, {Croton}, {Helly}, {Peacock}, {Cole},
  {Thomas}, {Couchman}, {Evrard}, {Colberg}, and {Pearce}]{2005Natur.435..629S}
V.~{Springel}, S.~D.~M. {White}, A.~{Jenkins}, C.~S. {Frenk}, N.~{Yoshida},
  L.~{Gao}, J.~{Navarro}, R.~{Thacker}, D.~{Croton}, J.~{Helly}, J.~A.
  {Peacock}, S.~{Cole}, P.~{Thomas}, H.~{Couchman}, A.~{Evrard}, J.~{Colberg},
  and F.~{Pearce}.
\newblock {Simulations of the formation, evolution and clustering of galaxies
  and quasars}.
\newblock {\em \nat}, 435:\penalty0 629--636, June 2005.

\bibitem[{Teyssier}(2002)]{2002A&A...385..337T}
R.~{Teyssier}.
\newblock Cosmological hydrodynamics with adaptive mesh refinement. a new high
  resolution code called ramses.
\newblock {\em \aap}, 385:\penalty0 337--364, April 2002.

\bibitem[{Teyssier} et~al.(2007){Teyssier}, {Pires}, {Aubert}, {Pichon},
{Prunet}, {Amara}, {Benabed}, {Colombi}, {Refregier}, {Starck}]{Teyssier2007}
R.~{Teyssier}, S.~{Pires}, D.~{Aubert}, C.~{Pichon}, S.~{Prunet},
A.~{Amara}, K.~{Benabed}, S.~{Colombi}, A.~{Refregier},
J.-L.~{Starck}.
\newblock {Full-Sky Weak Lensing Simulation with 70 Billion Particles}.
\newblock submitted to {\em \aap}, 2007.

\bibitem[{van de Weygaert}(1996)]{1996ASPC...94...49V}
R.~{van de Weygaert}.
\newblock {Tidal Fields and Structure Formation}.
\newblock In P.~{Coles}, V.~{Martinez}, and M.-J. {Pons-Borderia}, editors,
  {\em Mapping, Measuring, and Modelling the Universe}, volume~94 of {\em
  Astronomical Society of the Pacific Conference Series}, pages 49--+, 1996.

\bibitem[{Wadsley} et~al.(2004){Wadsley}, {Stadel}, and
  {Quinn}]{2004NewA....9..137W}
J.~W. {Wadsley}, J.~{Stadel}, and T.~{Quinn}.
\newblock {Gasoline: a flexible, parallel implementation of TreeSPH}.
\newblock {\em New Astronomy}, 9:\penalty0 137--158, February 2004.

\bibitem[{Weinberg} et~al.(2002){Weinberg}, {Hernquist}, and
  {Katz}]{2002ApJ...571...15W}
D.~H. {Weinberg}, L.~{Hernquist}, and N.~{Katz}.
\newblock {High-Redshift Galaxies in Cold Dark Matter Models}.
\newblock {\em \apj}, 571:\penalty0 15--29, May 2002.

\end{thebibliography}

\end{document}